\documentclass{acmsmall-ec13}
\usepackage{fancyhdr}
\usepackage{bbm}
\usepackage{amsfonts}
\usepackage{graphicx}
\usepackage{dcolumn}
\usepackage{booktabs}
\usepackage{caption}
\usepackage[caption=false]{subfig} 
\usepackage[all]{xy}
\usepackage{float}

\usepackage[ruled]{algorithm2e}
\renewcommand{\algorithmcfname}{ALGORITHM}
\SetAlFnt{\small}
\SetAlCapFnt{\small}
\SetAlCapNameFnt{\small}
\SetAlCapHSkip{0pt}
\IncMargin{-\parindent}

\newcommand{\indicator}[1]{\mathbbm{1}\left( {#1} \right) }

\begin{document}
\lhead{} 
\rhead{} 
\lfoot{} 
\cfoot{} 
\rfoot{} 
\chead{\bfseries Proceedings Article}  


\title{Selection Effects in Online Sharing: Consequences for Peer Adoption}
\author{
SEAN J. TAYLOR
\affil{NYU Stern, Facebook$^{\dag}$}
EYTAN BAKSHY
\affil{Facebook}
SINAN ARAL
\affil{NYU Stern}
}

\begin{abstract}

Most models of social contagion take peer exposure to be a corollary of adoption, yet in many settings, the visibility of one's adoption behavior happens through a separate decision process.  In online systems, product designers can define how peer exposure mechanisms work: adoption behaviors can be shared in a passive, automatic fashion, or occur through explicit, active sharing.  The consequences of these mechanisms are of substantial practical and theoretical interest: passive sharing may increase total peer exposure but active sharing may expose higher quality products to peers who are more likely to adopt.

We examine selection effects in online sharing through a large-scale field experiment on Facebook that randomizes whether or not adopters share Offers (coupons) in a passive manner. We derive and estimate a joint discrete choice model of adopters' sharing decisions and their peers' adoption decisions. 
Our results show that active sharing enables a selection effect that exposes peers who are more likely to adopt than the population exposed under passive sharing.
We decompose the selection effect into two distinct mechanisms: active sharers expose peers to higher quality products, and the peers they share with are more likely to adopt independently of product quality. Simulation results show that the user-level mechanism comprises the bulk of the selection effect. The study's findings are among the first to address downstream peer effects induced by online sharing mechanisms, and can inform design in settings where a surplus of sharing could be viewed as costly. 

\end{abstract}




\begin{bottomstuff}
$\dag$ This research was conducted while the author was visiting Facebook.

Author's addresses: Sean J. Taylor staylor@stern.nyu.edu, Eytan Bakshy eytan@fb.com, Sinan Aral sinan@stern.nyu.edu
\end{bottomstuff}

\maketitle
\thispagestyle{fancy}

\section{Introduction}

Standard models of social contagion consider adoption decisions of agents in the presence of social signals, but often take peer exposure to be a consequence of adoption~\cite{bass1969ms,granovetter1978threshold,jackson2007diffusion,schelling1973hockey}. This is natural for many situations where adoption creates a persistent signal that peers can observe; if an individual buys a car, she will find it difficult to prevent her peers from knowing about her adoption.  While theory tends to conflate adoption and exposure, they reflect substantive design decisions in practice.  In online settings, developers and marketers may seek to increase their ``virality'' by providing encouragements and incentives to spread their product or message to others. 

The decision to adopt and peer exposure can range from perfectly correlated to completely independent.  Applications which implement passive sharing automatically broadcast users' behaviors to their peers~\cite{aral2011creating}.  Similarly, Liking a Page on Facebook induces publicly visible connection that persists over time~\cite{bakshy2012social}.  However, in other settings (e.g. browsing the Web), individuals must actively share their behaviors.

In this paper, we examine the interaction between propogation and adoption. Exactly how an individual's adoption decision is linked to peer exposure can vary depending on the medium and product or message being spread. The relationship can be modulated by providing adopters with encouragements to share\footnote{We use the term \emph{share} because it broadly covers a range of diffusion phenomenon online, but we use it to address any choice that increases the visibility of an individual's adoption decision to others.  In some settings, this action can be seen as an endorsement or a form of conspicuous consumption \cite{veblen2005theory}.} that may impact the diffusion process. Developers can engineer features which encourage sharing or make peer exposure a more reliable consequence of product adoption or use\footnote{Passive sharing, where an action is automatically made visible to an individuals' peers, can be seen as the strongest possible encouragement to share, so that all users are equally as likely to share upon adoption.}, fulfilling the role that prestige and attractiveness might play in offline goods and services~\cite{veblen2005theory}. 
In order to predict the consequences of these strategies for product diffusion, it is important to understand how the nature of a user's decision to share may have an impact on her peers' decisions to adopt.

We implement a large-scale field experiment on Facebook to measure the effect of sharing interface on peer adoption, randomizing whether users of a coupon product share their redemption behavior in an active or passive manner.  Our results show that when adoption is perfectly linked with sharing, exposed peers are less likely to adopt the product in turn. The adoption effect we observe could be explained by either selection of influential adopters and/or susceptible peers (\emph{dyad selection}), or by selection of higher quality products. To disentangle these effects, we derive a discrete choice model of the adoption and sharing process and use Bayesian estimation techniques to fit the model to our experimental data. We find strong evidence for dyad selection -- peers of users who share in the active sharing condition are more likely to adopt the products their peers share.  We also find evidence for selection on product quality, but the effect on overall adoption outcomes is small. The major downstream effects of sharing regime are dominated by the selection of dyads.

We proceed as follows.  In the first section we review relevant literature on information diffusion in social networks and show how our work contributes to it.  In the following section, we derive an econometric model linking sharing with peer adoption in Section~\ref{sec:theory}.  We then describe our empirical context and experimental design in Section~\ref{sec:experiment}.  We summarize our results and estimate the selection effects using the experimental data and model in Section~\ref{sec:results}.  Finally, we conclude with a discussion of our results and their implications in Section~\ref{sec:discussion}.

\section{Related Work}\label{sec:related}

Online social networks allow users to articulate their relationships to people, companies, and products, and enable studies of diffusion processes \emph{in vivo} across consumer behaviors, including product recommendations~\cite{leskovec2007dynamics},
the adoption of social applications~\cite{aral2009distinguishing,wei2010diffusion}, and link re-sharing~\cite{bakshy2011everyone,goel2012structure}.

Recent studies are also beginning to analyze mechanisms of information transmission and their causal interpretations.  Since individuals form relationships with similar others~\cite{mcpherson2001birds}, network autocorrelation does not necessarily imply that an individuals influence their peers' behaviors \cite{hill2006network,aral2009distinguishing}.  This problem is exacerbated when the assumed exposure model omits backdoor paths which could plausibly account for the correlations \cite{shalizi2010homophily}. Even given perfect observability of the network process and abundant behavioral data, latent homophily or confounding factors could drive the assortativity in peer outcomes.

One of the most promising approaches to address these confounds in diffusion studies is the use of randomized field experiments~\cite{aral2011creating,aral2012identifying,bakshy2012social,bakshy2012diffusion}. However, experiments thus far either compare overall effects via different channels of influence, or focus on the direct effects of social signals on individual behavior through a single mechanism.  For example, \citeN{aral2011creating} showed that active personalized messaging is more effective in encouraging adoption per message,
while passive sharing generates greater total peer adoption in the network. However, it is not clear whether these differences in adoption rates are due to greater persuasiveness of the message format, differences in delivery\footnote{In the context of the authors' study, active messages were always received by the alter, whereas passive sharing could be aggregated or filtered, and therefore may not be as salient or seen by the alter.}, or selection effects.  Our work elucidates the mechanism for this selection effect (e.g. selecting peers who are more likely to adopt) by considering effects via single channel of communication and identical information content.

An intriguing aspect of word-of-mouth diffusion is the idea that social networks could be cleverly leveraged to increase the spread of desirable behaviors~\cite{hill2006network}.  Much of the research in this area approaches the problem via mathematical models that are analyzed through proofs on a given graph structure or through simulation~\cite{kempe2003maximizing,aral2011engineering,chierichetti2012schedule}.  Influence maximization has thus far been studied through models that do not account network autocorrelation in susceptibility, or make a distinction between adoption rates and the decision to share.  Part of the goal of our work is to shed light on how sharing decisions can affect downstream adoption to further the development of such models.

\section{Theory: selection mechanisms in product diffusion}\label{sec:theory}

Word-of-mouth (WOM) marketing or so-called ``viral'' diffusion is a repeated process of users adopting products and transmitting information about those products to their peers. While most studies focus on economic aspects of the \emph{adoption decisions} or patterns of diffusion over the network, we examine \emph{propagation decisions} and their selection effects on peer adoption.

\subsection{Selection mechanisms}
We posit that the conditions under which an adopter decides to share a product has a substantive impact the adoption rate of her peers.  To see why, consider sharing to be a binary choice for all adopters and consider two extreme cases: passive (automatic) sharing and active (selective) sharing.  In the former case, every adopter shares their adoption decision with all their peers, and sharing can perform no selective role.  In the latter, perhaps only a small fraction of adopters share.  This may happen, for instance, if sharing bears a cost -- such the time to send an email or the initiative to bring up a product in conversation.  To isolate selection, one can think of peers of sharers receiving identical signals indicating that their peer has adopted the product, e.g. a structured signal such a Like, a ``+1'', or a check-in.

In a world where peers receive homogeneous signals -- such as the aforementioned online social signals -- differences in adoption rates among peers of adopters in the two sharing regimes must be caused by selection of some combination of sharer-peer-product groups from different sample populations.  There are at least three possible selection effects that could alter peer adoption decisions when the sharing decision changes:

\begin{enumerate}
\item Adopter selection: Adopters who share are more influential than non-sharers.
\item Peer selection: Peers of sharers are more interested adopting the product than peers of non-sharers.
\item Product selection: Individuals share better products and decline to share worse ones.
\end{enumerate}

All three selection mechanisms could be present if an individual has some pro-social or financial motivation to share products that her peers are more likely to adopt.  In this case, she may use her private information about her peers' preferences to decide whether or not to share.
Product selection may be more salient if users gain utility from portraying an association with prestigious brands or if users receive disutility from creating associations that embarrass them \cite{akerlof2000economics}.

The difference between adopter and peer selection is subtle and worthy of further discussion.  For example, a so-called ``influential'' may cause many peers to adopt merely because her peers are easily influenced~\cite{watts2007influentials}.
It is clearly difficult to define a distinction between adopter and peer selection, and more difficult still to econometrically identify the difference between them.\footnote{One strategy for identifying adopter-specific influence requires repeated observation of different adopters sending persuasive messages to members of the same population of peers. Since each adopter may have different peers, it is difficult to design an experiment where peers are sampled from the same population.}
Accordingly, we will use the term \emph{dyad selection} to refer to selection of adopter-peer pairs where peer adoption is more likely, regardless of whether it is due to characteristics of the adopter, their peer, or the relationship between them. \emph{Product selection} will be used to refer to any selection effects which are observable between products.

\subsection{Model}\label{sec:model}
We now develop an economic model that formalizes our hypotheses and the tradeoffs we wish to describe.

\subsubsection{Choice model}

We index newly adopting individuals by $i = 1,\ldots,N$ and products by $k =
1,\ldots,K$.  We model the probability an individual makes her
adoption visible ($s_{ik} = 1$)  or not ($s_{ik} = 0$) as a piecewise
function depending on whether she is in the active sharing condition, indicated by $z_i = 1$. 

\begin{displaymath}
   s_{ik} = \left\{
     \begin{array}{lr}
       1 & : z_i = 0 \\
       \indicator{\mu_k + \epsilon_{ik} \geq 0} & : z_i = 1
     \end{array}
   \right.
\end{displaymath} 

Here $\mathbbm{1}$ is the indicator function and $\mu_k \sim \mbox{Normal}(\alpha, \sigma_{\mu}^2)$ is the population average random utility for sharing product $k$. The unobserved term $\epsilon_{ik} \sim \mbox{Normal}(0, \sigma_{\epsilon}^2)$ represents individual-specific factors that increase the utility of $i$ sharing $k$ with her peers.  This could include, for example, any utility she expects from sharing based on private information about her peers' preferences.

Individuals in the passive sharing condition live in a simple world: if they adopt, their behavior is visible to their peers.  Subjects in the
active sharing condition will share if the product itself is
``shareable'' enough -- $\mu_k$ is sufficiently high -- or if her
idiosyncratic utility $\epsilon_{ik}$ from sharing is high enough.
The first obvious result from this model is that sharing for those in
the active sharing condition is always less than in the passive
condition. We assume $\mu_k$ and $\epsilon_{ik}$ are independent random variables. In an empirical context, $\mu_k$ would be identified by variation in sharing rates between products.

Let $y_{ik} \in \{0, 1\}$ represent the decision of a peer of $i$,
the subject's peer, to adopt product $k$ after observing $i$'s behavior.  We model the peer adoption decision as the following discrete choice:

\begin{displaymath} 
  y_{ik} = \indicator{\lambda_k + \nu_{ik} \geq 0} .
\end{displaymath}

In this equation $\lambda_k \sim \mbox{Normal}(\gamma,
\sigma_{\lambda}^2)$ is the product utility from
adoption.  $\nu_{ik} \sim \mbox{Normal}(0, \sigma_{\nu}^2)$ is the
unobserved utility from adoption that $i$'s peers receive for $k$.
Note that the peer either receives a signal about the product or they
do not, they have no information about the individual's treatment status $z_i$\footnote{It is important to note that our experimental design isolates a pure selection effect because the messages received by peers are the same regardless of the treatment status of sender.}. Peers are only eligible to make this decision if $s_{ik} = 1$. 
We will now differentiate between product and dyad selection mechanisms, correlations across different utility components which link an
individual's sharing decision with her peers' adoption decisions. These mechanisms will only function if the individual's sharing decision
occurs in the active sharing regime.

\subsubsection{Selection on products}
Sharing can act as a selection mechanism on product quality.  Assume that $\mu_k$ and $\lambda_k$ are correlated random variables with correlation coefficient $\rho$, which would be the case if the latent product quality, e.g. value, provides both sharing and adoption utilities:
\begin{displaymath}
\left[
    \begin{array}{c}
      \mu_k \\
      \lambda_k
    \end{array}
  \right]
\sim \mbox{Normal} \left(
  \left[
    \begin{array}{c}
      \alpha \\
      \gamma
    \end{array}
  \right],
  \left[
    \begin{array}{cc}
      \sigma_{\mu}^2  &  \rho \hspace{.03cm}  \sigma_{\mu} \sigma_{\lambda} \\
      \rho \hspace{.03cm} \sigma_{\mu} \sigma_{\lambda} & \sigma_{\lambda}^2  \\
    \end{array}
  \right]
\right) .
\end{displaymath}

The correlation coefficient $\rho$ may be interepreted as a measure
of product selection and we hypothesize that $\rho > 0$: products that are more likely to be shared are also more likely to be adopted, independently of dyadic preferences.  The consequence of this hypothesis is that exposed peers decide whether to adopt products with a higher mean utility for adoption. To see how, let $\phi$ and $\Phi$ be the standard normal density and distribution functions respectively, and $\mathbb{E}$ be the expectation operator.  When adopters in the active sharing world share, the conditional distribution of $\lambda_k$ for her peers' adoption decisions has a higher expected value:

\begin{equation} \label{eq:lambdabias}
\mathbb{E} \left[ \lambda_k | \mu_k \geq - \epsilon_{ik} \right]
= 
\gamma + 
\sigma_{\lambda} \hspace{.03cm}  \rho \hspace{.03cm} 
\frac{\phi \left( -\frac{\epsilon_{ik}}{\sigma_{\mu}} \right ) }
     {1 - \Phi \left( -\frac{\epsilon_{ik}}{\sigma_{\mu}} \right) } 
     \geq \gamma
.
\end{equation}

In an empirical context, $\rho$ can be identified by the correlation in sharing and adoption rates between products.

\subsubsection{Selection on dyads}\label{sec:dyad_selection}
An individual who actively shares may be more likely to
generate adoptions from her peers than those who do not.  This could be because the
individual is
influential or her peers are susceptible to this particular product
(e.g. the individual recommends a product to peers who are likely adopters).\footnote{This type of selection mechanism through correlated unobservable variables is a contribution of \citeN{heckman1979sample}, which uses the example of researchers only observing the market wages of individuals who choose to enter the labor market.  Here we only observe the adoption decisions of users whose peers chose to make their adoptions visible.}

As in the \citeN{heckman1979sample} model of sample selection bias, we will assume that our unobserved utility components $\epsilon_{ik}$ and $\nu_{ik}$ are distributed bivariate normal with correlation coefficient $\psi$:
\begin{displaymath}
  \left[
    \begin{array}{c}
      \epsilon_{ik} \\
      \nu_{ik}
    \end{array}
  \right]
\sim \mbox{Normal} \left(
  \left[
    \begin{array}{c}
      0 \\
      0
    \end{array}
  \right],
  \left[
    \begin{array}{cc}
      \sigma_{\epsilon}^2    &  \psi \hspace{.03cm}  \sigma_{\epsilon} \sigma_{\nu} \\
      \psi \hspace{.03cm} \sigma_{\epsilon} \sigma_{\nu}  &  \sigma_{\nu}^2        \\
    \end{array}
  \right]
\right) 
.
\end{displaymath}

$\psi$ is our measure of dyad selection and we hypothesize that $\psi > 0$: an individual is more likely to
share when her peers are more likely to adopt the product, independent of the product quality.  As in the product selection mechanism, the effect is driven
by an increase in the mean of the distribution of the peer's idiosyncratic utility conditional on individual choosing to share:

\begin{equation} \label{eq:nubias}
\mathbb{E} \left[ \nu_{ik} | \epsilon_{ik} \geq - \mu_k \right]
= 
\sigma_{\nu} \psi 
\frac{\phi \left( -\frac{\mu_k}{\sigma_{\nu}} \right) }
     {1 - \Phi \left( -\frac{\mu_k}{\sigma_{\nu}} \right) }
\geq 0
.
\end{equation}

When $\psi > 0$, those who actively share will have peers who are more interested in
adopting the product.  In other words, passive sharing causes the individual to spread the product even when she knows her peers may not be likely to adopt.

If $\rho$ is identified through repeated observations of sharing and adoption behavior of products, then the correlation in individual specific
utilities $\psi$ can be identified by exogenously assigning the individual to
active and passive sharing interfaces.  By assumption, peers of
adopters in the passive condition have $\mathbb{E} [\nu_{ik}] =
0$, while peers of adopters in the active condition have idiosyncratic adoption utilities with conditional expectation, $\mathbb{E} [\nu_{ik} | \epsilon_{ik} > - \mu_k] > 0$.  This difference, and therefore $\psi$, can be identified by variation in adoption rates within products and across peers of adopters in randomly assigned sharing interfaces.

If either of our two selection hypotheses are confirmed (i.e. $\rho >
0$ or $\psi > 0$), peers of adopters who passively share will have a
lower probability of adoption than peers of adopters who actively share.

\section{Experiment}\label{sec:experiment}

We conducted a field experiment on Facebook to compare selection mechanisms present in active and passive sharing regimes using Facebook Offers, a marketing product that allows businesses to share discounts with customers by posting an Offers to their Facebook Pages.  Offers are similar to coupons or discounts available through sites like Groupon or LivingSocial.  When a user claims (adopts) an Offer, she receives an email which must either be shown at the businesses' physical location to get the discount, or can be used to receive a discount in an online store.  Simultaneously, those who share passively\footnote{This is the default behavior for many activities on Facebook such as Liking a Page.} share their claim activity with their peers (friends).

\begin{figure*}[h]
\begin{center}
\centering
\subfloat[]{
\hspace{5pt}
\includegraphics[width=0.37\textwidth]{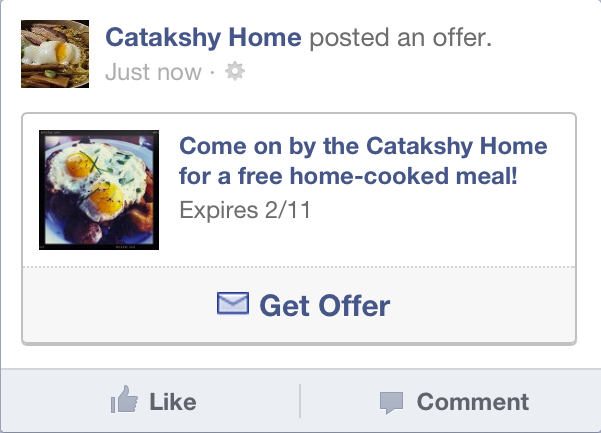}
\hspace{5pt}
}
\subfloat[]{
\hspace{5pt}
\includegraphics[width=0.37\textwidth]{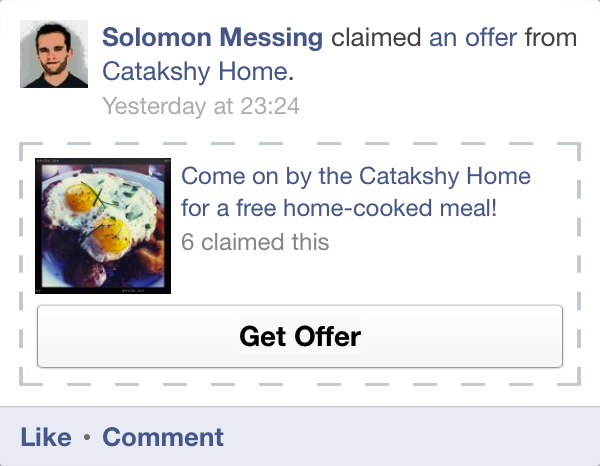}
\hspace{5pt}
} \\
\subfloat[]{
\includegraphics[width=0.6\textwidth]{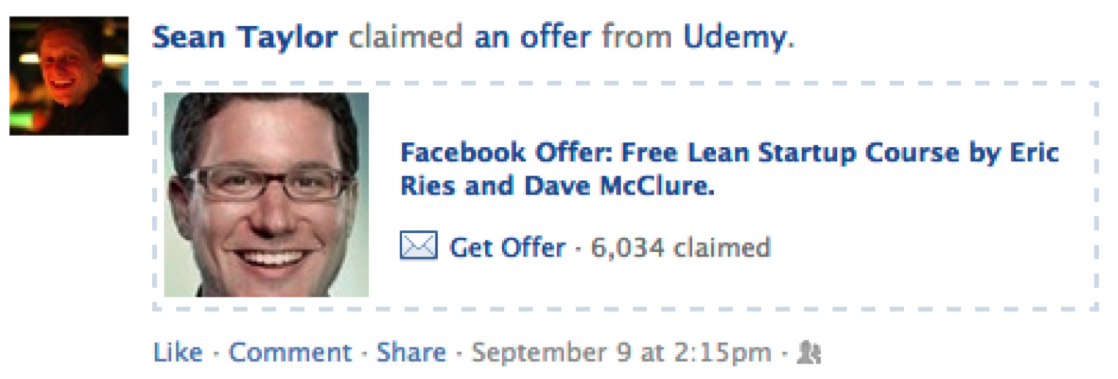}
}
\caption{A story for (a) a Page posting a new Offer on a mobile web browser and (b) a friend claiming (adopting) an Offer on the Facebook iPhone application (c) a friend claiming an Offer on the dekstop interface.}
\label{fig:claim-story}
\end{center}
\end{figure*}

Offers are distributed via Facebook's News Feed.  The News Feed is the primary means for users to consume \emph{stories} about friends' activities, such as status updates from friends, or from \emph{Pages}, which represent celebrities, businesses, and other organizations.  Thus, there are two ways that a user may receive an Offer.  First, the subscribers of a Page receive stories directly from the Page presenting the Offer (Figure~\ref{fig:claim-story}a).  Second, a user may be exposed to the Offer via a friend whose action was made visible after adopting it (Figure~\ref{fig:claim-story}b,c).   These two modes of diffusion correspond to having a single ``big seed'' (or broadcast node)~\cite{watts2007viral}  which initially spreads the Offer, after which point cascading effects may occur.

Our empirical context provides several advantages over other settings.  First, we can observe the diffusion of a large sample of comparable units, so our analyses do not suffer from survivor bias (i.e. we observe even the unsuccessful cascades).  Second, the behaviors we study (claiming Offers) provide users with valuable incentives, are low cost, and are expressly intended by marketers to achieve widespread distribution.  Third, many Offers receive substantial distribution and many adoptions, so we can observe many distinct users interact with the same Offer, which is crucial to our identification strategy.  Finally, we can plausibly observe almost all interactions between users and Offers because very little Offer transmission occurs outside of Facebook.

\subsection{Experimental design}

\begin{figure*}[t]
\begin{center}
\centering
\subfloat[]{
\hspace{5pt}
\includegraphics[width=0.37\textwidth]{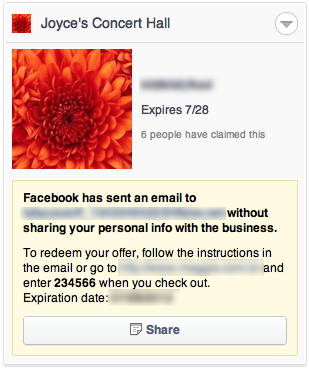}
\hspace{5pt}
}
\subfloat[]{
\hspace{5pt}
\includegraphics[width=0.37\textwidth]{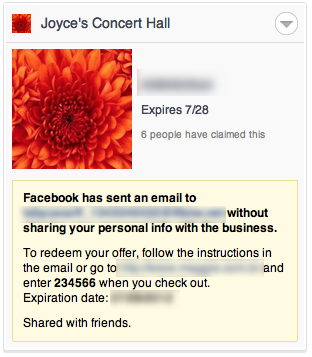}
\hspace{5pt}
}
\caption{Mobile interface presented to subjects after adoption for the (a) active and (b) passive sharing conditions.}
\label{fig:mobile-treatments}
\end{center}
\end{figure*}

1.2 million users were randomly assigned to one of two experimental conditions -- the active or passive sharing conditions -- with equal probability at the time of adoption.  That is, after subjects claimed an Offer (adopted) on a mobile device, they would either share their Offer redemption passively (Figure~\ref{fig:mobile-treatments}a), or were given a button that prompted the user to share their claim action with others (Figure~\ref{fig:mobile-treatments}b).  For each Offer, we record an impression event each time a user sees the Offer in their News Feed (Figure~\ref{fig:claim-story})
and if she claims (adopts) the Offer.  We also record whether she shared the Offer after adopting.   In the following analyses, we use this data and consider Offers that were claimed by at least 25 users during a two month period in 2012.

We examine downstream effects of the sharing interface by measuring the subsequent behavior of peers who were exposed to subjects' adoption activity.  It is important to note that peers who see the activity of subjects who share under the passive sharing condition are different than those who share in the active condition. This introduces a selection effect that shapes the population of exposed peers, and this effect is what we intend to measure.

\subsection{Interference}\label{sec:interference}
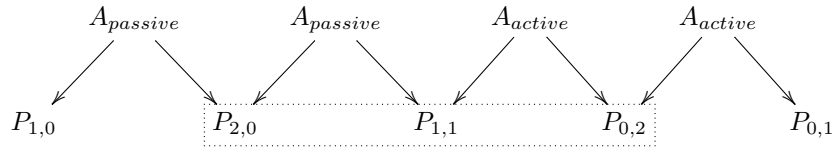
\begin{figure*}[h]
\begin{center}
\centering
\begin{displaymath}
  \xymatrix @C-1.5pc{
  & A_{passive} \ar[dl] \ar[dr] & &
  A_{passive} \ar[dl] \ar[dr] & &
  A_{active} \ar[dl] \ar[dr] & &
  A_{active} \ar[dl] \ar[dr] &  \\
  P_{1,0} & & P_{2,0} & & P_{1,1} & & P_{0,2} & & P_{0,1} 
  \save "2,3"."2,7"*[F.]\frm{} \restore
  }
\end{displaymath}
\caption{An illustration of potential interference patterns in our
  experiment. The subscripts for each exposed peer
  ($P$) denotes the number of adopters ($A$) in the passive and active
  condition who had exposed them to an Offer. Our analysis only considers peers of type $P_{0,1}$ and
  $P_{1,0}$, omitting those exposed via more than one adopter
  (e.g. peers in the dotted box) in order to isolate selection effects. These two types constitute approximately 90\% of exposed peers.}
\label{fig:exposure}
\end{center}
\end{figure*}

The experimental treatment -- a change in the mobile sharing interface -- is applied to adopters, but we measure the adoption outcomes of their peers. This approach can lead to interference if peers are exposed through multiple adopters (Figure~\ref{fig:exposure}), 
and is problematic for two reasons.  First, the status of a peer is no longer well defined if she is exposed by subjects in different conditions.
Second, even if a peer is exposed through multiple adopters with the same treatment, she may not be comparable to a user who is exposed through only one. Multiply-exposed individuals may have higher adoption rates due to increased homophily, multiple simultaneous social cues~\cite{bakshy2012social} or multiple exposures over time \cite{centola2010}.

Because passive sharing may increase the number of multiply-exposed peers, interference can confound our ability to identify the selection effects we wish to estimate.  Therefore our analysis only considers the peers who are exposed via a single adopter's sharing action\footnote{While we find effects from multiple exposures to be interesting, modeling these processes is beyond the scope of this paper.}. This preserves the vast majority (approximately 90\%) of exposed peers while simplifying interpretation of the results.

\section{Results}\label{sec:results}

We present our results through descriptive analysis and modeling.  The first subsection provides a basic overview of the experimental data.  In the following two subsections, we present results from reduced-form models which examine subjects' sharing decisions and peers' adoption decisions separately.  We focus on separating variation in sharing and adoption outcomes into variation in Offer-specific effects and idiosyncratic user effects. 
In the fourth subsection, we present estimates from the joint decision model introduced in Section~\ref{sec:model} to link the two models in a coherent system which can identify the correlation parameters we are interested in, allowing us to distinguish between product and dyad selection effects.

\subsection{Descriptive statistics}

Table~\ref{tab:ego-summary} shows summary results from the direct effect of the experiment. Approximately the same number of subjects were exposed to each sharing interface, and subjects in each condition were exposed via approximately the same number of distinct Offers. While all users in the passive condition shared, approximately one in five subjects in the active sharing  condition shared the Offer with their peers.

\begin{table}[t]
\small
\begin{center}
\begin{tabular}{lrr}
\toprule
& Active & Passive
  \\[0.2mm] \toprule \\[-2.25mm]
Subjects & 577,933 & 573,113 \\ 
  Distinct Offers & 23,102 & 23,251 \\ 
  Proportion shared & 0.23 & 1.00
    \\[0.5mm]  \bottomrule
\end{tabular}
\caption{Summary of statistics for direct effects on
  subjects' sharing behavior in the active and passive sharing conditions.}
\label{tab:ego-summary}
\end{center}
\end{table}

\begin{table}[t]
\small
\begin{center}
\begin{tabular}{lrr}
\toprule
& Active & Passive
  \\[0.2mm] \toprule \\[-2.25mm]
Mean friends exposed & 59.17 & 66.62 \\ 
  Median friends exposed & 41 & 46 \\ 
  Number of adoptions & 20,591 & 87,686 \\ 
  Adoption rate & 0.0050 & 0.0045 \\ 
  Adoptions per subject & 0.036 & 0.153 \\ 
  Adoptions per sharer & 0.157 & 0.154
    \\[0.5mm]  \bottomrule
\end{tabular}
\caption{Summary statistics for subjects' exposed peers in the active and
  passive sharing conditions.}
\label{tab:alter-summary}
\end{center}
\end{table}

\begin{figure}[h]
\centering
\includegraphics[width=0.65\textwidth]{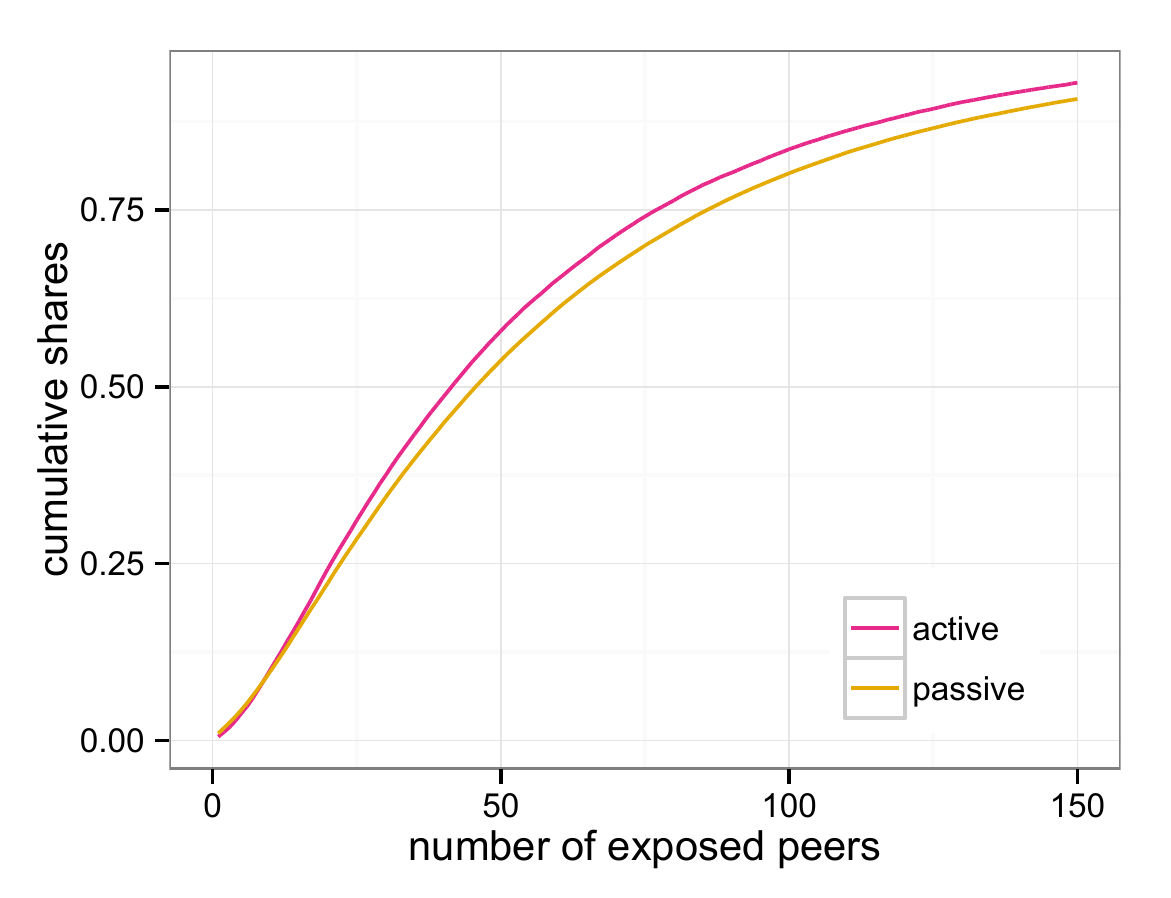}
\caption{The empirical cumulative distribution function for the number of exposed peers for each sharing subject by treatment condition. For clarity in comparison, the x-axis is truncated at the 90\% percentile of the distribution. The empirical distributions show sharers in the passive condition usually expose more of their peers.}
\label{fig:outdegree}
\end{figure}

After a subject shares an Offer, a story showing that the subject claimed the Offer was eligible to appear in her peers' News Feeds. Table~\ref{tab:alter-summary} provides descriptive statistics about how many peers were exposed to this story, as well as their subsequent adoption decisions.\footnote{Recall from Section~\ref{sec:interference} that we only consider the subpopulation of exposed peers who were exposed the Offer via a single friend.} 
The mean and median number of exposed peers is slightly higher for sharing subjects in the passive sharing condition compared to those in the active condition. Figure~\ref{fig:outdegree} shows the distribution of the number of exposed peers by treatment condition.  Here, we can see that active sharing shifts the distribution toward individuals who expose fewer friends to Offers. This effect is likely caused by selection on users who have fewer or or less active peers. The result is fewer social exposures to the Offers from both less sharing as well as smaller number of exposures per sharing user.

Peers who are reached via active sharing are more responsive on average with about a 10\% increase in the probability of adoption\footnote{All confidence intervals reported in this section use the multiway bootstrap~\cite{owen2012bootstrapping} with 500 replicates clustered by subjects and Offers.  This bootstrap is expected to be accurate even in situations where treatment effects are expected to vary with both subjects and items~\cite{bakshy2013bootstrap}.} (95\% confidence interval $[1.063, 1.134]$). However, the low sharing rate for subjects in the active condition means that it is about $4.3$ times more effective to enable passive sharing as measured by aggregate peer adoptions.

\begin{figure}[h]
\centering
\includegraphics[width=0.99\textwidth]{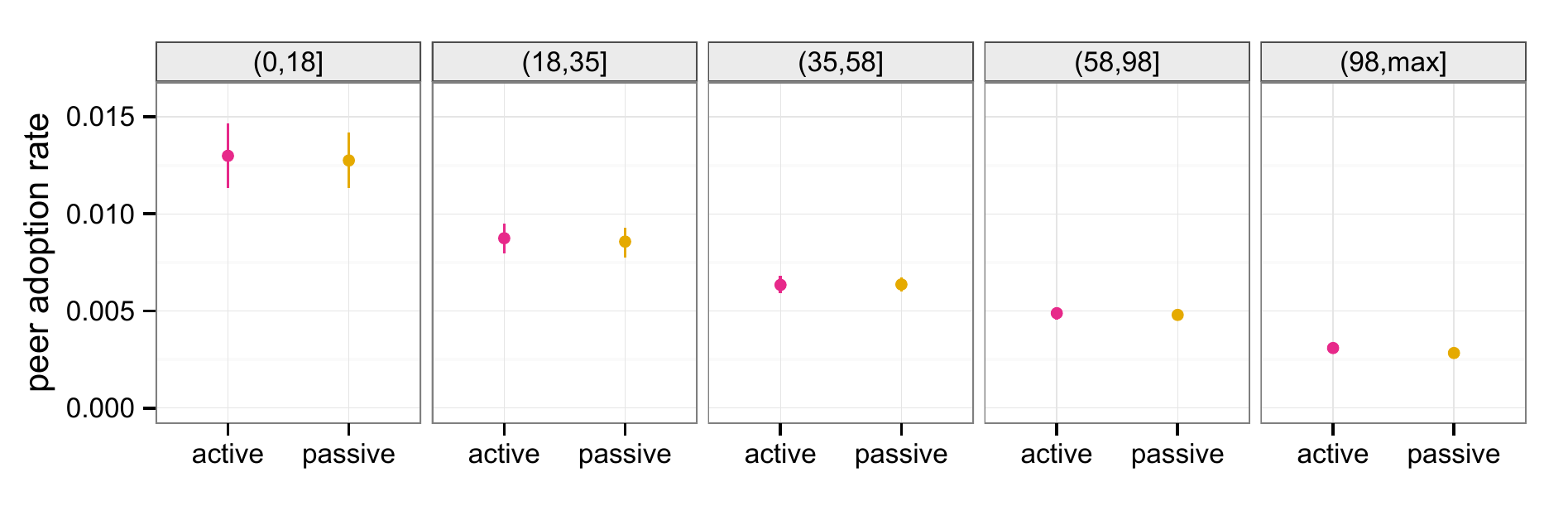}
\caption{Average adoption rate of sharing subjects' peers, broken down by quintiles of the total number of exposed peers. Error bars show the 95\% confidence intervals.
}
\label{fig:adoption-outdegree}
\end{figure}

Figure~\ref{fig:adoption-outdegree} provides an intuition for one of the mechanisms underlying selection effects in sharing decisions. Adoption rate of peers varies according to whether the the alter was exposed as a member of a large group of exposed peers of the individual or a small group. Furthermore, the variability in adoption rate is greater for those who expose fewer peers.

\subsection{Modeling variation in sharing behaviors}

We first report the results for those in the active sharing condition.  The share rate is approximately 23\% and is very precisely estimated.  We are interested in the extent to which Offer-level effects are driving sharing decisions by users.  If there is little variability between Offers and most of the variation occurs at the dyad level, then it will not be possible for the product selection mechanism to function.  To see why, assume there is no Offer-level variation in share rates.  Then Offer characteristics do not affect sharing at all and the set of shared Offers will be sampled from the same distribution as the unshared Offers.

We estimate a random effects probit model,

\begin{displaymath}
   Pr(s_{ik} = 1 | \mu_k) = \Phi(\mu_k + \epsilon_{ik})
,
\end{displaymath} 

in order to estimate $\sigma_{\mu}^2$.  To identify the parameters in the probit model, we let $\sigma_{\epsilon}^2 = 1$.  Table~\ref{tab:share} contains the model parameter estimates we obtain.  The estimated intraclass correlation coefficient is $\frac{\sigma_{\mu}^2}{\sigma_{\mu}^2 + \sigma_{\epsilon}^2} = 0.064$, indicating that the Offer-level random effects do not explain much additional variance in the sharing model.  This implies that the product selection mechanism is likely to be weak.

\begin{table}[]
\begin{center}
\begin{tabular}{lD{.}{.}{3}}
\toprule
Parameter & \multicolumn{1}{c}{Estimate} \\
\toprule
$\alpha$            & -0.812^{*}\\
                  &  (0.005)    \\
\midrule
$\sigma_{\mu}$        &   0.262     \\
\midrule
Log-likelihood    &  \multicolumn{1}{r}{-303,070} \\
Groups            &   \multicolumn{1}{r}{23,102}  \\
N                 &  \multicolumn{1}{r}{577,933}  \\
\bottomrule
\end{tabular}
\caption{Maximum likelihood parameter estimates for probit regression predicting share rate with random effects at the Offer level. The estimated mean of $\mu_k$ is $\alpha$, which is an estimate of average sharing utility.  The variance of the random effects at the Offer level $\mu_k$ is small compared to the total variance. *~denotes significance at the 0.001 level.}
\label{tab:share}
\end{center}
\end{table}

\subsection{Effect of passive sharing on downstream adoption}

In this section, we estimate an average treatment effect of active sharing on peer adoption rates.  For each subject who shares -- either because they were in the passive sharing condition or they chose to share in the active sharing condition -- we measure two aggregate outcomes of the subjects' peers: exposures and adoptions.

We define the number of peer exposures $n_{ik}$ for user $i$ and Offer $k$ to be the number of unique peers who saw a story in News Feed about the subject claiming the Offer.  We only count exposures which were unique, meaning that the alter must not have seen the Offer through any other user's adoption.  We count a peer as exposed just once regardless of how many impressions of claim story the user may have been served in her News Feed.

We define number of peer adoptions, $a_{ik} = \sum_{j=1}^{n_{ik}} y_{ijk}$, as a count of the number of peer exposures which generated an adoption.
We assume that $n_{ik}$ is exogenous, since it depends on the subject and her peers' characteristics and Facebook usage behavior.
Recall that $z_i$ represents the exogenous (experimental) manipulation of the subject's sharing interface and is equal to $1$ in for users in the active sharing condition.

\begin{displaymath}
   Pr(a_{ik} = L | n_{ik}) 
= \mbox{Binomial} \left( n_{ik}, \Phi(\beta z_i + \lambda_k + \nu_{ik}) \right)
,
\end{displaymath} 

where $\beta$ represents the average selection effect on the subject's peers.  As in the last section, we ignore the correlations between the unobserved parameters.  We report parameter estimates for the regression model in Table~\ref{tab:claim}.  The coefficient $\beta$, measures all selection effects, is positive and significant, and therefore confirms our hypothesis that active sharing will increase the probability that an subject's peers will adopt the product.  
The magnitude of $\beta$ corresponds to about a 7\%  marginal increase in the relative risk of adoption for peers of users who share in the passive condition (95\% confidence interval: $[1.050, 1.089]$).

\begin{table}[]
\begin{center}
\begin{tabular}{lD{.}{.}{3}}
\toprule
Parameter & \multicolumn{1}{c}{Estimate} \\
\toprule
$\gamma$          & -2.742^{*}\\
                  &  (0.004)    \\
$\beta$           &  0.022^{*}\\
                  &  (0.003)    \\
\midrule
$ \sigma_{\lambda} $ &   0.172     \\
\midrule
Log-likelihood    & \multicolumn{1}{r}{-151,521} \\
Groups            &   \multicolumn{1}{r}{25,726}     \\
N                 &  \multicolumn{1}{r}{702,090}     \\
\bottomrule
\end{tabular}
\caption{Maximum likelihood parameter estimates for binomial regression predicting the number of adopting alters with random effects at the Offer level. $\gamma$ is the mean of the random effect $\lambda_k$, while $\beta$ represents a reduced form measure of total selection effect. *~denotes significance at the 0.001 level.}
\label{tab:claim}
\end{center}
\end{table}

As in the sharing model, we have assumed $\sigma_{\nu}^2 = 1$ in order to identify the other parameters.  We can compute the intraclass correlation coefficient for adoption, $\frac{\sigma_{\lambda}^2}{\sigma_{\lambda}^2 + \sigma_{\nu}^2} = 0.029$.  This is low, indicating that product quality does not explain a large amount of the variance in adoption outcomes.  

\subsection{Joint decision model}

The structural model we specified in Section~\ref{sec:model} unifies the regression models in the previous two sections by accounting for the correlations between the unobserved parameters.  We assumed a correlation structure which accommodates two mechanisms for individual's sharing decision to impact her peers' adoption decisions, obviating the need for the $\beta$ parameter in the adoption model.  Estimating the joint model allows us to understand the relative contribution of each of the selection effects.

Our setup is similar to the simultaneous discrete-choice models with interdependent preferences considered in \citeN{yang2006estimating}, motivating a similar estimation procedure using Bayesian methods.  Bayesian estimation is ideal for this setting because it allows us to flexibly perform inference on the correlation parameters $\rho$ and $\psi$ at the cost of parametric assumptions.

We use non-informative priors on all parameters and run Markov-chain Monte Carlo simulations to estimate their posterior distributions given the observed data.  Due to the scale of our data, we used an efficient Hamiltonian Monte Carlo sampler \cite{hoffman2012} and computed our results using a state-of-the-art Bayesian model compiler \cite{stan-software:2013}.  We simulated three Markov chains for 2,000 iterations, discarding the first 1,000 iterations for ``burn-in.''  We then used the last 1,000 draws for estimation. We evaluated convergence by computing a potential scale reduction factor for each estimated parameter in the model \cite{gelman1992inference}.

\begin{table}[ht]
\begin{center}
\begin{tabular}{llrrrr}
  \toprule
Parameter & & Mean & 2.5\% & Median & 97.5\% \\ 
  \toprule
  Mean product sharing utility & $\alpha$ 
  & -0.813 & -0.821 & -0.813 & -0.804 \\ 
  Mean product adoption utility & $\gamma$ 
  & -2.739 & -2.748 & -2.739 & -2.730 \\ 
  Std. dev. of product sharing utility & $\sigma_{\mu}$ 
  & 0.267 & 0.258 & 0.267 & 0.279 \\ 
  Std. dev. of product adoption utility & $\sigma_{\lambda}$ 
  & 0.172 & 0.165 & 0.173 & 0.179 \\ 
  Product-level correlation coefficient & $\rho$ 
  & 0.174 & 0.102 & 0.176 & 0.238 \\ 
  Dyad-selection correlation coefficient & $\psi$ 
  & 0.025 & 0.019 & 0.025 & 0.032 \\ 
   \bottomrule
\end{tabular}
\caption{HMC posterior mean, median, and 95\% credible interval estimates for the parameters of the joint structural model of sharing and adoption specified in Section~\ref{sec:model}. Estimates for $\rho$ and $\psi$ are positive and significant, providing evidence for both product- and dyad-selection effects.}
\label{tab:joint}
\end{center}
\end{table}

We estimate three main types of parameters in the model (Table~\ref{tab:joint}).  First, there are mean sharing and adoption utilities, $\alpha$ and $\gamma$, which rationalize the average rates of sharing and adoption.  Second, there are correlations between the unobserved utilities at the product level, $\rho$, and at the dyad level, $\psi$.  Third, we estimate the standard deviations of the zero-mean product-level utilities, $\sigma_{\mu}$ and $\sigma_{\lambda}$.

Note that as in the regression models, $\sigma_{\epsilon}$ and $\sigma_{\nu}$ are fixed at $1$ in order to identify the other parameters of the model.  This is a typical assumption in this modeling situation where we have no absolute utility scale.

We find evidence for both types of selection effects that we hypothesized, $\rho > 0$ and $\psi > 0$, indicating that in the active sharing interface, users who shared shared Offers which were more likely to be adopted and seen by peers who were more likely to adopt them.\footnote{Recall from Section~\ref{sec:dyad_selection} that one cannot distinguish between adopter and peer selection mechanisms in our setting.  Part of the positive correlation $\psi$ may be explained by influential users' higher propensity to share.}

It is worth pointing out that our estimates of $\alpha$, $\gamma$, and the variances of the random effects $\sigma_{\mu}^2$ and $\sigma_{\lambda}^2$ are extremely close to the reduced-form models of the previous section.  We have essentially replaced $\beta$ with structural correlation parameters which allow us to distinguish between two mechanisms.  But which of these mechanisms is more important?

Our estimate for $\rho$ is substantially larger than our estimate for $\psi$, which could be interpreted as evidence for the relative importance of product selection effects.  However, we must consider that the distributions of the Offer-level and dyad-level effects are different in scale.  This warrants further analysis of the interaction between correlations and effect scales.

With posterior distributions for parameters in hand, we can use our model to decompose the treatment effect into product and peer selection through simulation.  Recall the relative risk of peer adoption for active versus passive sharing had 95\% confidence interval $[1.063, 1.134]$, which is measured by

\begin{displaymath}
\mbox{RR} = \frac{Pr(y_{ik} = 1 | z_i = 1, s_{ik} = 1)}{Pr(y_{ik} = 1 | z_i = 0)} .
\end{displaymath}

The effect of $z_i$ works through two exhaustive mechanisms. First it changes $\rho$ from $0$ to our estimate $\hat{\rho} > 0$, enabling selection on product quality. Second, it changes $\psi$ from $0$ to our estimate $\hat{\psi} > 0$, enabling selection on dyads. We can simulate relative risk under counterfactuals scenarios where only one of the two mechanisms is enabled:

\begin{displaymath}
\mbox{RR}_{product} = \frac{Pr(y_{ik} = 1 | s_{ik} = 1, \rho = \hat{\rho}, \psi = 0)}{Pr(y_{ik} = 1 | s_{ik} = 0)} ;
\end{displaymath}

\begin{displaymath}
\mbox{RR}_{dyad} = \frac{Pr(y_{ik} = 1 | \rho = 0, \psi = \hat{\psi})}{Pr(y_{ik} = 1 | s_{ik} = 0}) .
\end{displaymath}

To compute these counterfactual relative risks, we simulate sharing behavior and subsequent adoption rates by drawing from our posterior parameter distributions. For $\mbox{RR}_{product}$ we set $\psi = 0$ and then draw $(\epsilon_{ik}, \nu_{ik})$ pairs as independent random variables. For $\mbox{RR}_{dyad}$ we set $\rho = 0$ and then draw independent $(\mu_{k}, \lambda_{k})$ pairs. We then compute empirical relative risks over $500$ generated sample populations and compute means and 95\% confidence intervals for selection effects under each counterfactual scenario.

\begin{figure}[h]
\centering
\includegraphics[width=0.65\textwidth]{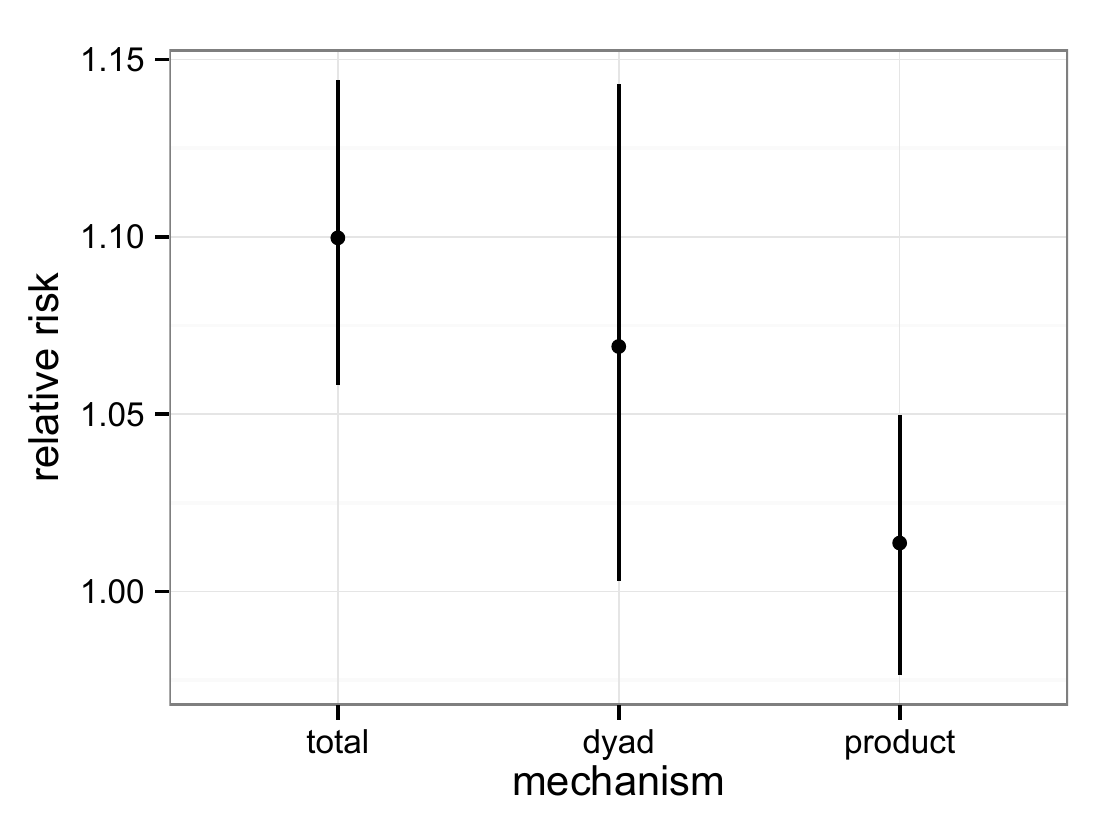}
\caption{Simulated relative risks of adoption for active versus passive sharing. 500 iterations were used, each with a set of parameters drawn from the estimated posterior distributions. Dyad selection comprises the bulk of the selection effect in most cases and the mechanisms are complementary. The confidence intervals on the total relative risk are larger than those we report earlier because they incorporate model uncertainty.}
\label{fig:simresults}
\end{figure}

The results of this procedure are shown in Figure~\ref{fig:simresults}. We can see that disabling the product selection, leaving dyad selection only, retains most of the total selection effect in our simulations. In comparison, the product selection effect is weaker. Thus, despite the high correlation in Offer sharing and adoption utilities, their relatively low importance in the explaining the variance of these decisions limits the product selection effect.

\section{Discussion}\label{sec:discussion}

We have presented a theoretical result and supporting evidence that encouraging so-called ``virality'' decreases the efficiency of marketing messages in social networks.  Our study is the first to identify the interaction between adoption and propagation decisions.  This relationship is important because peers of users who choose to share, and the products they share, are potentially different than the peers of users and products shared by the general population of adopters.  Our results suggest that the decision to share enhances efficiency of diffusion by increasing the probability of adoption for downstream users.  Thus when users can choose to share, there are fewer wasted exposures generated in the diffusion process.

From a design perspective, our results show that while encouraging users to share their behaviors may increase the total number of adoptions, it can have negative consequences.  There exists a tradeoff for platform providers for whom distribution is a scarce resource or brands using costly incentive strategies to improve rates of peer exposure.


In our experimental setup, either active or passive sharing distributed adoption stories through an automated content ranking system, exposing a potentially large audience to identical messages.  In other settings, the audience and message resulting from an adopter's sharing decision may be more variable. Adopters may decide how many peers they share with, with whom they share, and what exactly they choose to say when they share.  It is possible that giving adopters tighter control over the outcome from sharing could yield stronger selection effects than we observed, resulting in higher adoption rates.

Our parameter estimates also seem to suggest a potential explanation for why campaigns rarely ``go viral''~\cite{bakshy2011everyone,goel2012structure}.  In order to propagate through a network, a product must be adopted and shared at a high rate.  In Facebook Offers, we found that product-level factors which predict adoption and sharing are only mildly correlated and explain only a small fraction of the variance in spreading behaviors.  It may simply be rare to find examples of products which contribute high levels of sharing and adoption utility to all users.

\subsection{Limitations and Future Work}

While we are able to plausibly observe users' interactions with respect to sharing and claiming Offers comprehensively, we are bound to investigate selection effects that occur via plausible changes to Facebook's existing delivery mechanisms.  For Facebook users, sharing means publishing content
to a specified audience -- often friends -- so that the content appears in friends' News Feeds.
Like face-to-face situations, the likelihood of receiving information via the Feed is determined by an individual's preferences and previous interactions.  Thus, it is possible that Facebook's feed ranking algorithm automatically plays some selective role in the diffusion of Offers.  It is possible that other platforms, especially those which do not use ranking, to exhibit even stronger sharing selection effects.



We used a randomized field experiment to estimate selection effects in the sharing process for single individuals and their peers.  Since selection effects may compound over several steps of the diffusion process, it is possible that individual-level effects may differ for subjects had the experiment randomized over Offers.  Furthermore, passive sharing is likely to increase the number of social signals an individual receives.  As discussed in Section~\ref{sec:interference}, multiply-exposed peers may behave differently, and we might also expect that overall macro effects would be different under other randomization schemes.  While our experiment is designed specifically to estimate peer effects not caused by multiple reinforcing signals, examining how these different effects interact would be of substantial interest from a policy perspective.

There are also a number of other opportunities for further exploration.  In our setting, adoption and sharing decisions were relatively costless for the subjects, requiring only a single touch on a mobile phone.  It would be interesting to see if these results apply for more costly settings where adoption comes at some expense.  Other types of encouragements to sharing could be explored, such as monetary incentives, which could generate smoother variation in the rate of sharing.  Finally, other peer outcomes, such as using the Offer in brick-and-mortar stores, are also of great interest.

\section{Acknowledgements}
We would like to thank Dean Eckles, Rohit Dhawan, and Cameron Marlow
for their feedback on this work.

\bibliographystyle{acmsmall}
\bibliography{references}

\begin{thebibliography}{}

\bibitem[\protect\citeauthoryear{Akerlof and Kranton}{Akerlof and
  Kranton}{2000}]{akerlof2000economics}
{\sc Akerlof, G.~A.} {\sc and} {\sc Kranton, R.~E.} 2000.
\newblock Economics and identity.
\newblock {\em The Quarterly Journal of Economics\/}~{\em 115,\/}~3, 715--753.

\bibitem[\protect\citeauthoryear{Aral, Muchnik, and Sundararajan}{Aral
  et~al\mbox{.}}{2009}]{aral2009distinguishing}
{\sc Aral, S.}, {\sc Muchnik, L.}, {\sc and} {\sc Sundararajan, A.} 2009.
\newblock {Distinguishing influence-based contagion from homophily-driven
  diffusion in dynamic networks}.
\newblock {\em Proceedings of the National Academy of Sciences\/}~{\em
  106,\/}~51, 21544.

\bibitem[\protect\citeauthoryear{Aral, Muchnik, and Sundararajan}{Aral
  et~al\mbox{.}}{2011}]{aral2011engineering}
{\sc Aral, S.}, {\sc Muchnik, L.}, {\sc and} {\sc Sundararajan, A.} 2011.
\newblock Engineering social contagions: Optimal network seeding and incentive
  strategies.
\newblock In {\em available at SSRN: http://ssrn. com/abstract}. Vol. 1770982.

\bibitem[\protect\citeauthoryear{Aral and Walker}{Aral and
  Walker}{2011}]{aral2011creating}
{\sc Aral, S.} {\sc and} {\sc Walker, D.} 2011.
\newblock Creating social contagion through viral product design: A randomized
  trial of peer influence in networks.
\newblock {\em Management Science\/}~{\em 57,\/}~9, 1623--1639.

\bibitem[\protect\citeauthoryear{Aral and Walker}{Aral and
  Walker}{2012}]{aral2012identifying}
{\sc Aral, S.} {\sc and} {\sc Walker, D.} 2012.
\newblock Identifying influential and susceptible members of social networks.
\newblock {\em Science\/}~{\em 337,\/}~6092, 337--341.

\bibitem[\protect\citeauthoryear{Bakshy and Eckles}{Bakshy and
  Eckles}{2013}]{bakshy2013bootstrap}
{\sc Bakshy, E.} {\sc and} {\sc Eckles, D.} 2013.
\newblock Uncertainty in online experiments with dependent data: {A}n
  evaluation of bootstrap methods.
\newblock {\em Arxiv preprint. http://arxiv.org/abs/1304.7406\/}.

\bibitem[\protect\citeauthoryear{Bakshy, Eckles, Yan, and Rosenn}{Bakshy
  et~al\mbox{.}}{2012a}]{bakshy2012social}
{\sc Bakshy, E.}, {\sc Eckles, D.}, {\sc Yan, R.}, {\sc and} {\sc Rosenn, I.}
  2012a.
\newblock Social influence in social advertising: {E}vidence from field
  experiments.
\newblock In {\em Proceedings of the 13th ACM Conference on Electronic
  Commerce}. ACM, 146--161.

\bibitem[\protect\citeauthoryear{Bakshy, Hofman, Mason, and Watts}{Bakshy
  et~al\mbox{.}}{2011}]{bakshy2011everyone}
{\sc Bakshy, E.}, {\sc Hofman, J.~M.}, {\sc Mason, W.~A.}, {\sc and} {\sc
  Watts, D.~J.} 2011.
\newblock Everyone's an influencer: {Q}uantifying influence on {T}witter.
\newblock In {\em Proceedings of the fourth ACM international conference on Web
  search and data mining}. WSDM '11. ACM, New York, NY, USA, 65--74.

\bibitem[\protect\citeauthoryear{Bakshy, Rosenn, Marlow, and Adamic}{Bakshy
  et~al\mbox{.}}{2012b}]{bakshy2012diffusion}
{\sc Bakshy, E.}, {\sc Rosenn, I.}, {\sc Marlow, C.}, {\sc and} {\sc Adamic,
  L.} 2012b.
\newblock The role of social networks in information diffusion.
\newblock In {\em Proceedings of the 21st international conference on World
  Wide Web}. WWW '12. ACM, New York, NY, USA, 519--528.

\bibitem[\protect\citeauthoryear{Bass}{Bass}{1969}]{bass1969ms}
{\sc Bass, F.~M.} 1969.
\newblock A new product growth for model consumer durables.
\newblock {\em Management Science\/}~{\em 15,\/}~5, 215--227.

\bibitem[\protect\citeauthoryear{Centola}{Centola}{2010}]{centola2010}
{\sc Centola, D.} 2010.
\newblock The spread of behavior in an online social network experiment.
\newblock {\em Science\/}~{\em 329,\/}~5996, 1194--1197.

\bibitem[\protect\citeauthoryear{Chierichetti, Kleinberg, and
  Panconesi}{Chierichetti et~al\mbox{.}}{2012}]{chierichetti2012schedule}
{\sc Chierichetti, F.}, {\sc Kleinberg, J.}, {\sc and} {\sc Panconesi, A.}
  2012.
\newblock How to schedule a cascade in an arbitrary graph.
\newblock In {\em Proceedings of the 13th ACM Conference on Electronic
  Commerce}. ACM, 355--368.

\bibitem[\protect\citeauthoryear{Gelman and Rubin}{Gelman and
  Rubin}{1992}]{gelman1992inference}
{\sc Gelman, A.} {\sc and} {\sc Rubin, D.} 1992.
\newblock Inference from iterative simulation using multiple sequences.
\newblock {\em Statistical science\/}~{\em 7,\/}~4, 457--472.

\bibitem[\protect\citeauthoryear{Goel, Watts, and Goldstein}{Goel
  et~al\mbox{.}}{2012}]{goel2012structure}
{\sc Goel, S.}, {\sc Watts, D.~J.}, {\sc and} {\sc Goldstein, D.~G.} 2012.
\newblock The structure of online diffusion networks.
\newblock In {\em Proceedings of the 13th ACM Conference on Electronic
  Commerce}. EC '12. ACM, New York, NY, USA, 623--638.

\bibitem[\protect\citeauthoryear{Granovetter}{Granovetter}{1978}]{granovetter1978threshold}
{\sc Granovetter, M.} 1978.
\newblock {Threshold models of collective behavior}.
\newblock {\em The American Journal of Sociology\/}~{\em 83,\/}~6, 1420--1443.

\bibitem[\protect\citeauthoryear{Heckman}{Heckman}{1979}]{heckman1979sample}
{\sc Heckman, J.} 1979.
\newblock {Sample selection bias as a specification error}.
\newblock {\em Econometrica: Journal of the econometric society\/}, 153--161.

\bibitem[\protect\citeauthoryear{Hill, Provost, and Volinsky}{Hill
  et~al\mbox{.}}{2006}]{hill2006network}
{\sc Hill, S.}, {\sc Provost, F.}, {\sc and} {\sc Volinsky, C.} 2006.
\newblock {Network-based marketing: Identifying likely adopters via consumer
  networks}.
\newblock {\em Statistical Science\/}~{\em 21,\/}~2, 256--276.

\bibitem[\protect\citeauthoryear{Hoffman and Gelman}{Hoffman and
  Gelman}{2012}]{hoffman2012}
{\sc Hoffman, M.~D.} {\sc and} {\sc Gelman, A.} 2012.
\newblock The no-{U}-turn sampler: Adaptively setting path lengths in
  {H}amiltonian {M}onte {C}arlo.
\newblock {\em Journal of Machine Learning Research\/}.

\bibitem[\protect\citeauthoryear{Jackson and Yariv}{Jackson and
  Yariv}{2007}]{jackson2007diffusion}
{\sc Jackson, M.} {\sc and} {\sc Yariv, L.} 2007.
\newblock {Diffusion of behavior and equilibrium properties in network games}.
\newblock {\em The American Economic Review\/}~{\em 97,\/}~2, 92--98.

\bibitem[\protect\citeauthoryear{Kempe, Kleinberg, and Tardos}{Kempe
  et~al\mbox{.}}{2003}]{kempe2003maximizing}
{\sc Kempe, D.}, {\sc Kleinberg, J.}, {\sc and} {\sc Tardos, {\'E}.} 2003.
\newblock {Maximizing the spread of influence through a social network}.
\newblock In {\em Proceedings of the ninth ACM SIGKDD international conference
  on Knowledge discovery and data mining}. ACM, 137--146.

\bibitem[\protect\citeauthoryear{Leskovec, Adamic, and Huberman}{Leskovec
  et~al\mbox{.}}{2007}]{leskovec2007dynamics}
{\sc Leskovec, J.}, {\sc Adamic, L.}, {\sc and} {\sc Huberman, B.} 2007.
\newblock The dynamics of viral marketing.
\newblock {\em ACM Transactions on the Web (TWEB)\/}~{\em 1,\/}~1, 5.

\bibitem[\protect\citeauthoryear{McPherson, Smith-Lovin, and Cook}{McPherson
  et~al\mbox{.}}{2001}]{mcpherson2001birds}
{\sc McPherson, M.}, {\sc Smith-Lovin, L.}, {\sc and} {\sc Cook, J.} 2001.
\newblock Birds of a feather: Homophily in social networks.
\newblock {\em Annual review of sociology\/}, 415--444.

\bibitem[\protect\citeauthoryear{Owen and Eckles}{Owen and
  Eckles}{2012}]{owen2012bootstrapping}
{\sc Owen, A.~B.} {\sc and} {\sc Eckles, D.} 2012.
\newblock Bootstrapping data arrays of arbitrary order.
\newblock {\em The Annals of Applied Statistics\/}~{\em 6,\/}~3, 895--927.

\bibitem[\protect\citeauthoryear{Schelling}{Schelling}{1973}]{schelling1973hockey}
{\sc Schelling, T.} 1973.
\newblock {Hockey helmets, concealed weapons, and daylight saving: A study of
  binary choices with externalities}.
\newblock {\em The Journal of Conflict Resolution\/}~{\em 17,\/}~3, 381--428.

\bibitem[\protect\citeauthoryear{Shalizi and Thomas}{Shalizi and
  Thomas}{2011}]{shalizi2010homophily}
{\sc Shalizi, C.~R.} {\sc and} {\sc Thomas, A.~C.} 2011.
\newblock Homophily and contagion are generically confounded in observational
  social network studies.
\newblock {\em Sociological Methods and Research\/}~{\em 27}, 211--239.

\bibitem[\protect\citeauthoryear{{Stan Development Team}}{{Stan Development
  Team}}{2013}]{stan-software:2013}
{\sc {Stan Development Team}}. 2013.
\newblock Stan: A c++ library for probability and sampling, version 1.1.

\bibitem[\protect\citeauthoryear{Veblen}{Veblen}{2005}]{veblen2005theory}
{\sc Veblen, T.} 2005.
\newblock {\em The theory of the leisure class; an economic study of
  institutions}.
\newblock Aakar Books.

\bibitem[\protect\citeauthoryear{Watts and Dodds}{Watts and
  Dodds}{2007}]{watts2007influentials}
{\sc Watts, D.} {\sc and} {\sc Dodds, P.} 2007.
\newblock Influentials, networks, and public opinion formation.
\newblock {\em Journal of consumer research\/}~{\em 34,\/}~4, 441--458.

\bibitem[\protect\citeauthoryear{Watts, Peretti, and Frumin}{Watts
  et~al\mbox{.}}{2007}]{watts2007viral}
{\sc Watts, D.~J.}, {\sc Peretti, J.}, {\sc and} {\sc Frumin, M.} 2007.
\newblock {\em Viral marketing for the real world}.
\newblock Harvard Business School Pub.

\bibitem[\protect\citeauthoryear{Wei, Yang, Adamic, de~Ara{\'u}jo, and
  Rekhi}{Wei et~al\mbox{.}}{2010}]{wei2010diffusion}
{\sc Wei, X.}, {\sc Yang, J.}, {\sc Adamic, L.}, {\sc de~Ara{\'u}jo, R.}, {\sc
  and} {\sc Rekhi, M.} 2010.
\newblock Diffusion dynamics of games on online social networks.
\newblock In {\em Proceedings of the 3rd conference on Online social networks}.
  USENIX Association, 2--2.

\bibitem[\protect\citeauthoryear{Yang, Narayan, and Assael}{Yang
  et~al\mbox{.}}{2006}]{yang2006estimating}
{\sc Yang, S.}, {\sc Narayan, V.}, {\sc and} {\sc Assael, H.} 2006.
\newblock {Estimating the Interdependence of Television Program Viewership
  Between Spouses: A Bayesian Simultaneous Equation Model}.
\newblock {\em Marketing Science\/}~{\em 25,\/}~4, 336.

\end{thebibliography}

\end{document}